\documentclass[3p,times,twocolumn]{elsarticle}
\usepackage{slashed}
\usepackage{amsmath}

\newcommand{\kg}{\boldsymbol{k}_\gamma} 
\newcommand{\kgp}{\boldsymbol{k}_{\gamma\perp}} 
\newcommand{\xp}{\boldsymbol{x}_\perp}
\newcommand{\up}{\boldsymbol{u}_\perp}
\newcommand{\yp}{\boldsymbol{y}_\perp}
\newcommand{\zp}{\boldsymbol{z}_\perp}

\newcommand{\rp}{\boldsymbol{r}_\perp}
\newcommand{\khp}{\boldsymbol{k}_{1\perp}} 
\newcommand{\kAp}{\boldsymbol{k}_{2\perp}} 
\newcommand{\kp}{\boldsymbol{k}_\perp}
\newcommand{\pp}{\boldsymbol{p}_\perp}
\newcommand{\Pp}{\boldsymbol{P}_\perp}
\newcommand{\qp}{\boldsymbol{q}_\perp}

\newcommand{\bp}{\boldsymbol{p}}
\newcommand{\bq}{\boldsymbol{q}}

\makeatletter
\DeclareRobustCommand{\cev}[1]{%
  \mathpalette\do@cev{#1}%
}
\newcommand{\do@cev}[2]{%
  \fix@cev{#1}{+}%
  \reflectbox{$\m@th#1\vec{\reflectbox{$\fix@cev{#1}{-}\m@th#1#2\fix@cev{#1}{+}$}}$}%
  \fix@cev{#1}{-}%
}
\newcommand{\fix@cev}[2]{%
  \ifx#1\displaystyle
    \mkern#23mu
  \else
    \ifx#1\textstyle
      \mkern#23mu
    \else
      \ifx#1\scriptstyle
        \mkern#22mu
      \else
        \mkern#22mu
      \fi
    \fi
  \fi
}

\makeatother

\usepackage{ecrc}

\volume{00}

\firstpage{1}

\journalname{Nuclear and Particle Physics Proceedings}

\runauth{}

\jid{nppp}

\jnltitlelogo{Nuclear and Particle Physics Proceedings}

\usepackage{amssymb}

\usepackage[figuresright]{rotating}

\begin{document}

\begin{frontmatter}



\dochead{}

\title{Photons from the Color Glass Condensate in p+A collisions}


\author[ad1,ad2]{Sanjin Beni\' c}
\author[ad2]{Kenji Fukushima}
\author[ad3]{Oscar Garcia-Montero}
\author[ad4]{Raju Venugopalan}

\address[ad1]{Physics Department, Faculty of Science, University of Zagreb, Zagreb 10000, Croatia}
\address[ad2]{Department of Physics, The University of Tokyo, 7-3-1 Hongo, Bunkyo-ku, Tokyo 113-0033, Japan}
\address[ad3]{Institut f\"{u}r Theoretische Physik, Universit\"{a}t Heidelberg, Philosophenweg 16, 69120 Heidelberg, Germany}
\address[ad4]{Physics Department, Brookhaven National Laboratory, Bldg.\ 510A, Upton, NY 11973, USA}

\begin{abstract}
We report on a first NLO computation of photon production in p+A collisions at collider energies within the Color Glass Condensate framework, significantly extending previous LO results.
At central rapidites, our result is the dominant contribution and probes multi-gluon correlators in nuclei.
At high photon momenta, the result is directly sensitive to the nuclear gluon distribution.
The NLO result contains two processes, the annihilation process and the process with $q\bar{q}$ pair and a photon in the final state. We provide a
numerical evaluation of the photon spectrum from the annihilation process.
\end{abstract}

\begin{keyword}
Color Glass Condensate \sep photon \sep nuclear collision

\end{keyword}

\end{frontmatter}


\section{Introduction}
\label{sec:intro}

Nuclei at high energies have a semi-classical description in terms of over-occupied gluons with fluctuations in the gluon density characterized by the saturation scale $Q_s$ as predicted by the Color Glass Condensate (CGC) effective theory \cite{Iancu:2003xm,JalilianMarian:2005jf,Gelis:2010nm}. Photon is a useful probe of CGC due to the absence of final state interactions. In this contribution we discuss the NLO calculation of photon production in p+A collisions.

\begin{figure}
\begin{flushleft}
\includegraphics[width=3cm,clip]{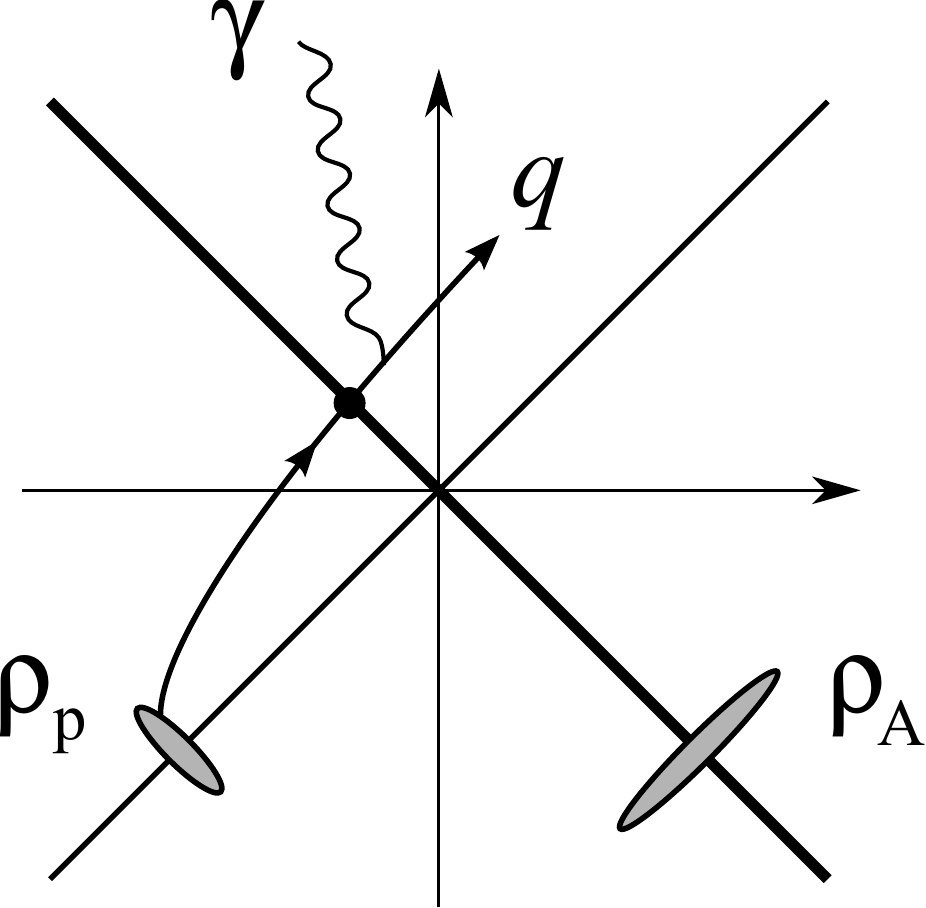}
\end{flushleft}
\includegraphics[width=3cm,clip]{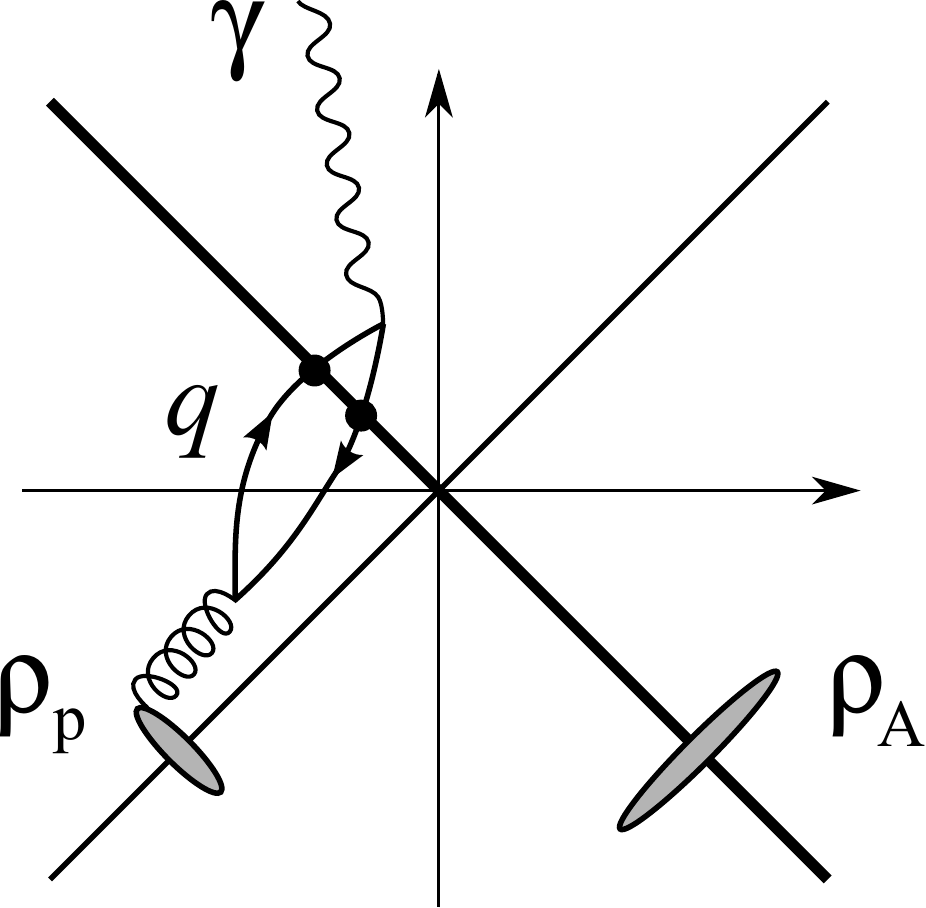}
\hspace{1.0cm}
\includegraphics[width=3cm,clip]{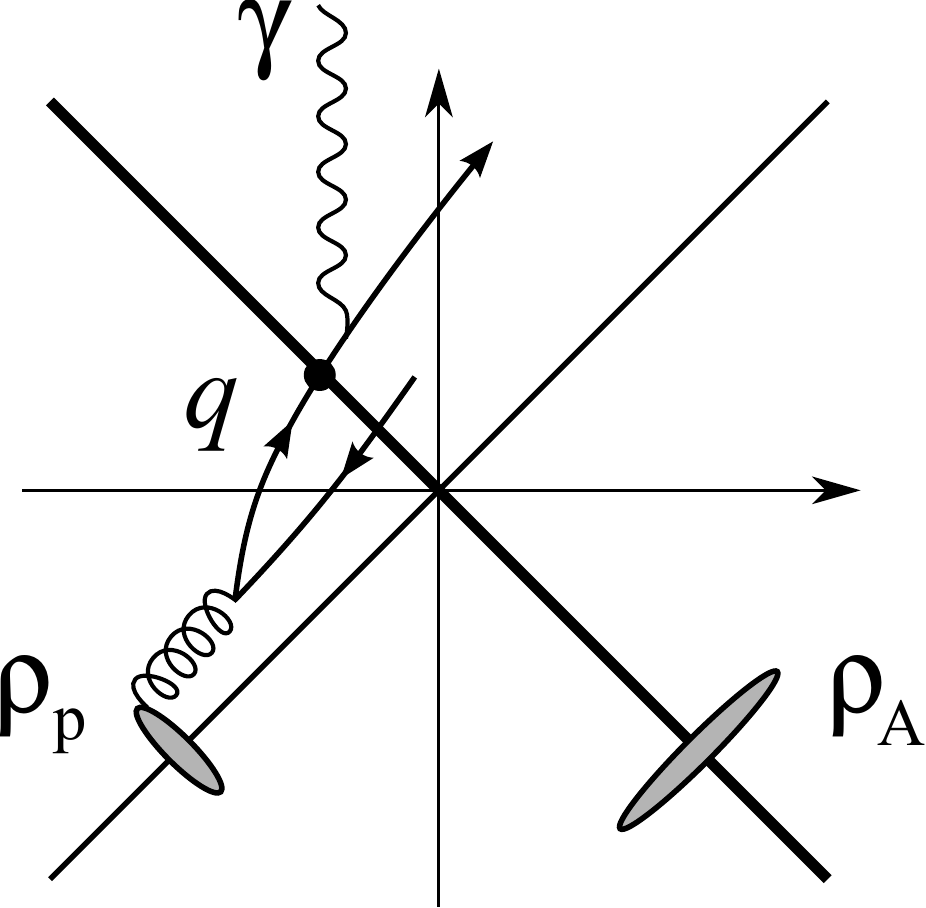}
\caption{Upper diagram is representative of LO contribution, while lower left (class II) and the lower right (class III) processes are representative of NLO contributions.}
\label{fig:diag}
\end{figure}

At the LO, photon is produced from valence quark bremsstrahlung (see Fig.~\ref{fig:diag}). The CGC result was calculated by Gelis and Jalilian-Marian in \cite{Gelis:2002ki} through which the photon spectrum becomes sensitive to the two-point gluon correlator of the nucleus. LO photon would apply at forward kinematics, where the valence quarks dominate the proton wave function. Further predictions in p+A include photon-hadron \cite{JalilianMarian:2012bd,Rezaeian:2012wa} and photon-jet \cite{Rezaeian:2016szi} correlations.

At large center of mass energy, or at central rapidities, the gluon component of the proton cannot be neglected. Working within the dilute-dense approximation, appropriate for the p+A kinematics, we investigate NLO processes with a gluon in place of the valence quark being emitted from the proton. In lower half of Fig.~\ref{fig:diag}, gluon emits a quark-antiquark pair. The pair either annihilates back to a photon (class II process \cite{Benic:2016yqt}) or goes to the final state (class III process \cite{Benic:2016uku}).
The inclusive rate in both cases is of the order $O(\alpha\alpha_s)$. To compare with valence quark bremsstrahlung we must multiply with the respective distribution functions for quarks $f_q$ and for gluons $f_g$ inside the proton. We consider the high energy regime where the new processes become dominant as $f_q \ll \alpha_s f_g$.

\section{Calculation of the diagrams}
\label{sec:calc}

We solve the classical Yang-Mills equations in the powers of the proton color density $\rho_p$ using the systematic dilute-dense approximation $\rho_p \ll \rho_A$ \cite{Blaizot:2004wu}, with $\rho_A$ being the nucleus density.
We denote the resulting $O(\rho_p^{0,1})$ gluon field as $\mathcal{A}^\mu_{(0,1)}(x)$.
In the $\mathcal{A}^+ = 0$ gauge the field of the nucleus alone, $\mathcal{A}^\mu_{(0)}$, is
\begin{equation}
 \mathcal{A}_{(0)}^\mu(x) = -g \delta^{\mu -}\delta(x^+)\frac{1}{\partial_\perp^2}\rho_A(\xp)\,.
\end{equation}
The field $\mathcal{A}^\mu_{(1)}$ is given in Fourier space in \cite{Gelis:2005pt,Fukushima:2008ya} as
\begin{align}
  &\mathcal{A}_{(1)}^\mu(q) =
    \frac{i g\, \delta^{\mu i} q_\perp^i}{(q^+ + i\epsilon)(q^- - i\epsilon)}\,
    \frac{\rho_p(\qp)}{\qp^2} + \frac{i g}{q^2+i q^+\epsilon}\nonumber\\
      &\times\int_{\khp}\int_{\xp} e^{i (\qp-\khp)\cdot\xp}\,
      C^\mu(q,\khp)\, U(\xp)\, \frac{\rho_p(\khp)}{\khp^2}\,,
\label{eq:A1l}
\end{align}
where $U(\xp)$ is the Wilson line with $\mathcal{A}^\mu_{(0)}$ in the adjoint representation
\begin{equation}
U(\xp) = \mathcal{P}_+ \exp\left[ig\int_{-\infty}^\infty d z^+ \mathcal{A}_{(0)}^-(z^+,\xp) \cdot T\right]\,.
\label{eq:wil}
\end{equation}
The quark propagator $S_{(0)}(x,y)$ is given from the Dirac equation in the $\mathcal{A}^\mu_{(0)}$ background \cite{McLerran:1998nk} as
\begin{align}
 & S_{(0)}(x,y) \equiv S_F(x-y) + i\theta(x^+)\theta(-y^+)\nonumber\\
 &\times\int_z\,
   \delta(z^+)\bigl[\tilde{U}(\zp)-1\bigr] S_F(x-z)\gamma^+ S_F(z-y)\nonumber\\ 
  & -i\theta(-x^+)\theta(y^+)\nonumber\\
  &\times\int_z\,
   \delta(z^+)\bigl[\tilde{U}^\dag(\zp)-1\bigr]S_F(x-z)\gamma^+ S_F(z-y)\,,
\label{eq:qprop}
\end{align}
with $\tilde{U}(\xp)$ being the Wilson line in the fundamental representation, that is, with the fundamental generators $t^a$, replacing the adjoint ones $T^a$ in Eq.~(\ref{eq:wil}).

The amplitude for the class II process with a single flavor of charge $e$ in the quark loop is
\begin{align}
 &\mathcal{M}_\lambda(\kg) = \frac{e g^2}{4\pi^3}\int_0^1 dx\int_{\yp \zp} \int_{\khp}\,
  e^{i(\kgp-\khp) \cdot [x\yp + (1-x)\zp]}\nonumber\\
 &\times \frac{\rho_p^a(\khp)}{k_{1\perp}^2} \, {\rm Tr}_c\bigl[
  \tilde{U}(\yp) t^a \tilde{U}^\dag(\zp)\bigr]\nonumber\\
    &\times \biggl\{\hat{u}_\lambda k_{1\perp}\Psi_1(\khp,\up,x)+ k_{1\lambda} \Psi_2(\khp,\up,x) \biggr\}\,,
\label{eq:ampII}
\end{align}
where $\kg$ is the photon momentum, $\up = \yp - \zp$, and $\Psi_{1,2}$ are given in \cite{Benic:2016yqt}.
We have checked that (\ref{eq:ampII}) satisfies the photon Ward identity.
To calculate the rate we square the amplitude and perform a color average over the sources.
The result can be summarized as \cite{Benic:2016yqt}
\begin{align}
&\frac{dN}{d^2 \kgp d\eta_{k_\gamma}}
   = \frac{\alpha_e \, \alpha_s}{16\pi^8}\frac{N_c}{N_c^2-1}
   \int_0^1 dx \, dx' \int_{\yp \yp' \zp \zp'} e^{i\kgp\cdot \rp} \nonumber\\
 &\times S(\yp,\zp,\yp',\zp')\int_{\khp} e^{-i\khp\cdot\rp}\varphi_p(\khp)\nonumber\\
&\times\bigl[\hat{\boldsymbol{u}}_\perp\cdot\hat{\boldsymbol{u}}_\perp' \Psi_1\Psi'^\ast_1
   + \Psi_2 \Psi'^\ast_2 + 2\hat{\boldsymbol{u}}_\perp\cdot\hat{\boldsymbol{k}}_{1\perp}\Psi_1\Psi'^\ast_2\bigr]\,.
\label{eq:rateII}
\end{align}
where 
\begin{align}
S(\yp,\zp,\yp',\zp') &\equiv \frac{1}{N_c}\biggl\langle \mathrm{Tr}_c \left[
  \tilde{U}\left(\yp\right) T_F^a \tilde{U}^\dag\left(\zp\right)\right]\\
  &\times \mathrm{Tr}_c \left[ \tilde{U}\left(\zp'\right) T_F^a \tilde{U}^\dag\left(\yp'\right)\right]\biggr\rangle\,,
\end{align}
is the inelastic quadrupole \cite{Dominguez:2011wm}.
$\varphi_p(\khp)$ is the unintegrated gluon distribution in the proton and $\rp=x\yp + (1-x)\zp - x'\yp' - (1-x')\zp'$.

We have expanded the Wilson lines to recover the perturbative calculation.
We have found that the $O(\rho_p\rho_A)$ contribution vanishes by charge conjugation. The lowest order non-vanishing contribution is $O(\rho_p\rho_A^2)$. We have explicitly checked that this is UV finite by gauge invariance.
Taking the zero quark mass limit is completely safe in our result as there are no collinear singularities.
Finally, we have checked that the term in the last line of Eq.~(\ref{eq:rateII}) has a well defined $\khp\to 0$ limit so that our result can be brought in a collinear factorized form on the proton side.

Turning to the class III process, in addition to the photon, the final state contains a quark with momentum $\bq$ and a antiquark with momentum $\bp$. The amplitude is
\begin{align}
&\mathcal{M}_\lambda(\kg,\bq,\bp) = -e g^2\int_{\kp \khp}\int_{\xp \yp} \frac{\rho_p^a(\khp)}{k_{1\perp}^2}\nonumber\\
&\times e^{i\kp\cdot\xp + i(\kgp + \qp + \pp - \kp - \khp)\cdot \yp}\nonumber\\
&\times\bar{u}(\bq)\epsilon_{\lambda\mu}(\kg)\biggl\{T^\mu_g(\khp)U(\xp)^{ba}t^b \nonumber\\
& + T_{q\bar{q}}^\mu(\kp,\khp)\tilde{U}(\xp)t^a \tilde{U}^\dag(\yp)\biggr\} v(\bp)\,.
\label{eq:ampIII2}
\end{align}
where the explicit expressions for the functions $T^\mu_g(\khp)$ and $T^\mu_{q\bar{q}}(\kp,\khp)$ can be found in \cite{Benic:2016yqt}. 
We have checked that (\ref{eq:ampIII2}) satisfies the photon Ward identity and also that the leading twist result satisfies the gluon Ward identities.
In general, the result for the amplitude is gauge dependent. We have made the calculation in the $\mathcal{A}^+ = 0$ and also in the $\partial_\mu \mathcal{A}^\mu = 0$ gauge \cite{Benic:2016uku,Garcia-Montero}. Interestingly, after some algebraic manipulation, the two results for the amplitude, within this particular gauge choices, become identical.

The total cross section for $\bar{q}q\gamma$ production is found as \cite{Benic:2016uku}
\begin{align}
&\frac{d\sigma}{d^2 \kgp d\eta_{k_\gamma} d^2 \qp d\eta_q d^2\pp d\eta_p} = \frac{\alpha_e\alpha_s^2}{256 \pi^8 C_F}\nonumber\\
&\times\int_{\khp\kAp} (2\pi)^2 \delta^{(2)}(\Pp - \khp - \kAp) \frac{\varphi_p(\khp)}{k_{1\perp}^2 k_{2\perp}^2}\Biggl\{\int_{\kp \kp'}\nonumber\\
&\times{\rm Tr}[(\slashed{q} + m)T^\mu_{q\bar{q}}(\khp,\kp)(m-\slashed{p}) \gamma^0 T^\dag_{q\bar{q}\mu}(\khp,\kp')\gamma^0]\nonumber\\
&\times \phi_A^{q\bar{q},q\bar{q}}(\kp,\kAp-\kp;\kp',\kAp-\kp')\nonumber\\
&+\int_{\kp}{\rm Tr}[(\slashed{q} + m)T^\mu_{q\bar{q}}(\khp,\kp)(m-\slashed{p}) \gamma^0 T_{g\mu}^\dag(\khp)\gamma^0]\nonumber\\
&\times\phi_A^{q\bar{q},g}(\kp,\kAp-\kp;\kAp) + {\rm h.\, c.}\nonumber\\
&+{\rm Tr}[(\slashed{q} + m)T^\mu_g(\khp)(m-\slashed{p}) \gamma^0 T_{g\mu}^\dag(\khp)\gamma^0]\nonumber\\
&\times\phi_A^{g,g}(\khp)
\Biggr\}\,,
\end{align}
where $\Pp = \kgp + \qp + \pp$.
The cross section depends on three different generalized unintegrated gluon distribution functions of the nucleus $\phi_A^{q\bar{q},q\bar{q}}$, $\phi_A^{q\bar{q},g}$ and $\phi_A^{g,g}$.
They correspond to different Wilson line correlators in the fundamental and in the adjoint representation identical to the ones found for $q\bar{q}$ production \cite{Blaizot:2004wv}.
The inclusive photon production is found by integrating over the quark and antiquark phase space.

Expanding the Wilson lines we find a contribution at $O(\rho_p \rho_A)$ (leading twist). This is the $k_\perp$-factorized photon production as it would be obtained by a pQCD computation.
Taking the collinear limit on the proton ($\khp\to 0$) and on the nucleus ($\kAp \to 0$) side, the inclusive cross section can be expressed as
\begin{align}
&\frac{d\sigma}{d^2 \kgp d\eta_{k_\gamma}} = \frac{1}{16}\int_0^\infty \frac{d q^+}{q^+}\frac{d p^+}{p^+}\int_{\qp\pp}(2\pi)^2 \delta^{(2)}(\Pp)\nonumber\\
&\times x_p f_{g,p}(x_p,Q^2)x_A f_{g,A}(x_A,Q^2)|\mathcal{M}_{gg\to q\bar{q}\gamma}|^2 \,.
\end{align}
From this result we get a clear relation between the photon spectrum and the gluon distribution in the nuclei.

In the limit of $\kgp\to 0$ the amplitude $\mathcal{M}_{gg\to q\bar{q}\gamma}$ factorizes in accordance to the soft photon theorem to a radiative part and the remaining, non-radiative part, responsible for $q\bar{q}$ production. The radiative part gives an infrared enhancement to the inclusive cross section as $1/\kgp^2$.

\section{Numerical results}
\label{sec:num}

\begin{figure}
\centering
\includegraphics[width=7cm,clip]{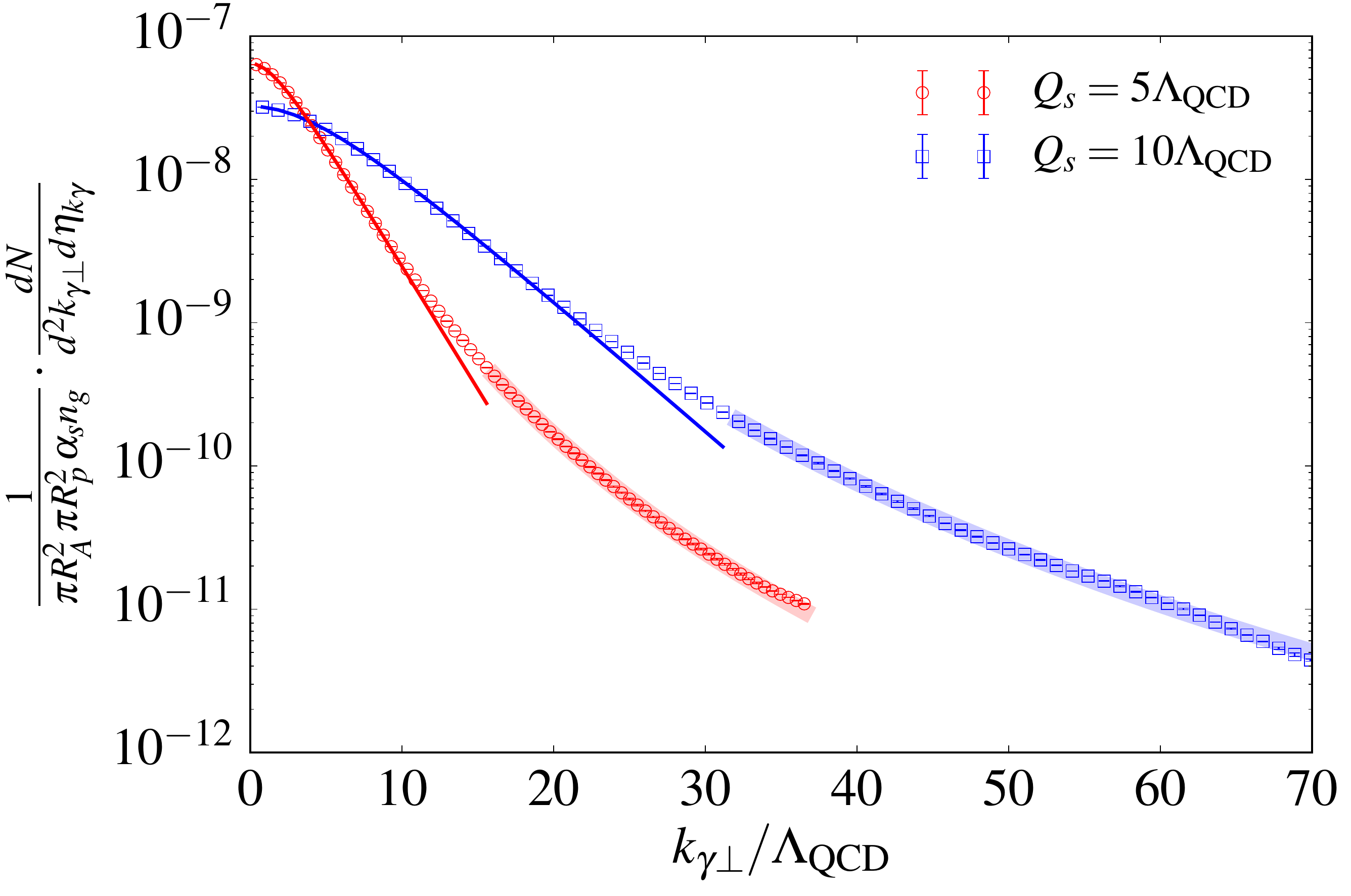}
\caption{Photon spectrum from the class II diagram. The results are shown for a single quark flavor and in the chiral limit.}
\label{fig:spec}
\end{figure}
Here we discuss the numerical computation of the class II process.
The general analytical expression (\ref{eq:rateII}) is valid for a single massive quark flavor.

The respective Wilson line correlator occurring in the class II process is referred to in the literature as the inelastic quadrupole \cite{Dominguez:2011wm}. Interestingly, within the large $N_c$ approximation, the inelastic quadrupole does not factorize. This means that one must handle a simultaneous integration over all the coordinates of the quadrupole.

For the numerical calculations we assume the case of a large nucleus, that is $A\gg 1$, in which case approximate translational invariance in the transverse plane holds. This  turns one transverse plane integration into a volume factor $\pi R_A^2$, where $R_A$ is the nuclear radii.
For the calculation of the inelastic quadrupole we use the McLerran Venugopalan model $\langle\rho^a_A(\xp) \rho^b_A(\yp)\rangle \equiv g^2 \delta^{ab} \mu_A^2  \delta^{(2)}(\xp - \yp)$ \cite{McLerran:1993ni,McLerran:1993ka,McLerran:1994vd}.
The saturation scale in the MV model is defined as
\begin{equation}
Q_s^2 \equiv \frac{N_c^2 - 1}{4 N_c}g^4 \mu_A^2\,.
\end{equation}
Finally, we determine $\varphi_p(\khp)$ from the MV model for the proton as $k_{1\perp}^2 \varphi_p(\khp) = \pi (N_c^2-1) g^4 \mu_p^2 \pi R_p^2$.

The numerical integration is optimized through a combination of the MISER Monte Carlo algorithm and a Quasi Discrete Hankel Transform algorithm, for details see \cite{Benic:2016uku}.
The computation is performed for two different values of the saturation scale $Q_s = 5 \Lambda_{\rm QCD}$ and $Q_s = 10 \Lambda_{\rm QCD}$. 
The numerical results for the transverse momentum spectrum of the photon is shown on Fig.~\ref{fig:spec}, where we factorize the transverse proton density fluctuation parameter as $n_g \equiv (N_c^2-1) g^4 \mu_p^2/(4 N_c)$.
For simplicity we have taken the chiral limit.

The parametrization of the soft part of the spectrum (up to $k_{\gamma\perp}\sim 2 Q_s$) is possible by an exponential as $\exp\left(-\sqrt{k_{\gamma\perp}^2 + (0.5 Q_s)^2}/0.5 Q_s\right)$, while semi-hard part with $k_{\gamma\perp}\gtrsim 2 Q_s$ is well fitted by a power law tail $\left(\log(k_{\gamma\perp}/Q_s)\right)^{1.5}/k_{\gamma\perp}^{5.6}$.
The thin lines on Fig.~\ref{fig:spec} are the exponential fit, while the thick lines are the power law fit.

\begin{figure}
\centering
\includegraphics[width=7cm,clip]{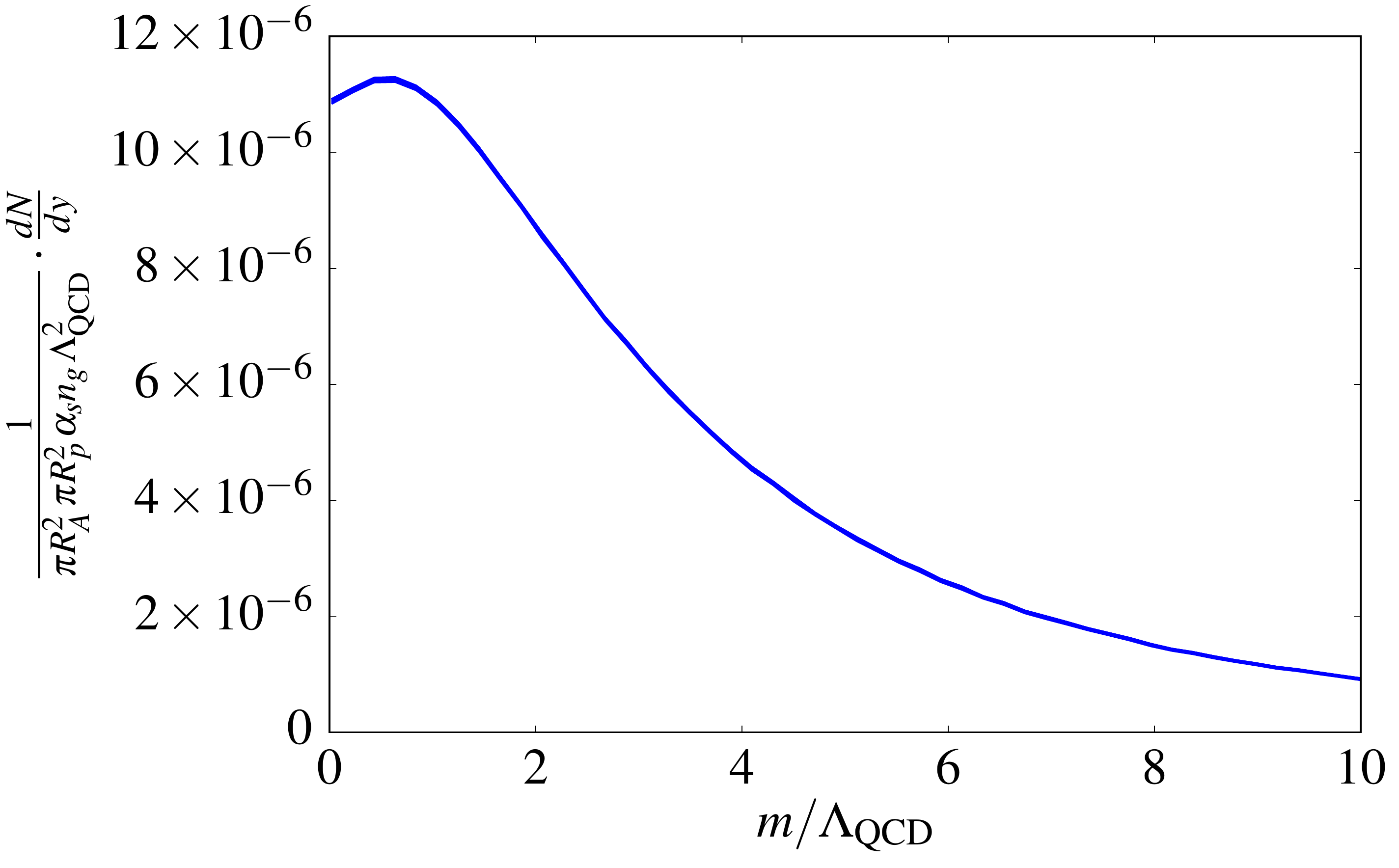}
\caption{Number of photons per unit rapidity as a function of the  quark mass $m$. Saturation scale is chosen as $Q_s = 10\Lambda_{\rm QCD}$.}
\label{fig:mass}
\end{figure}

On Fig.~\ref{fig:mass} we show the mass dependence of the number of photons per unit rapidity. Parametrization $(\log(m/\Lambda_{\rm QCD}))^{1.8}/m^{2.6}$ works well for $m\gtrsim 2\Lambda_{\rm QCD}$.

\section{Conclusions}
\label{sec:conc}

We have performed a detailed analytical calculation of NLO photon rates in p+A collisions.
The NLO photon rates probe multi-gluon correlations in the nuclei and as such substantiate and complement related investigations using hadron production \cite{Blaizot:2004wv,Fujii:2006ab,Fujii:2013gxa,Kang:2013hta,Ma:2015sia,Ma:2014mri}.
Our results will allow a direct study of the nuclear gluon distribution function, as well as the higher-twist corrections thereoff, in a systematic way.
Furthermore, they may help resolve an experimental puzzle \cite{Belogianni:2002ic} concerning the excess of soft photons in comparison to the predictions by the Low-Burnett-Kroll soft photon theorem.
Relative to hadrons, photons are free from hadronization uncertainties and dedicated photon measurements would go a long way towards exploring the physics of gluon saturation.

\section*{Acknowledgments}

S.~B.\ was supported by the European Union Seventh Framework Programme
(FP7 2007-2013) under grant agreement No.\ 291823, Marie Curie
FP7-PEOPLE-2011-COFUND NEWFELPRO Grant No.\ 48.
K.~F.\ was supported by MEXT-KAKENHI Grant No.\ 15H03652 and
15K13479.
R.~V.\ is supported under DOE Contract No. DE-SC0012704. 
This material is also based upon work supported by the U.S. Department of Energy, Office of Science, Office of Nuclear Physics, within the framework of the TMD Topical Collaboration. S.~B.\ acknowledges HZZO Grant No.\ 8799 at
Zagreb University for computational resources.




\nocite{*}
\bibliographystyle{elsarticle-num}
\bibliography{jos}

\begin{thebibliography}{10}
\expandafter\ifx\csname url\endcsname\relax
  \def\url#1{\texttt{#1}}\fi
\expandafter\ifx\csname urlprefix\endcsname\relax\def\urlprefix{URL }\fi
\expandafter\ifx\csname href\endcsname\relax
  \def\href#1#2{#2} \def\path#1{#1}\fi

\bibitem{Iancu:2003xm}
E.~Iancu, R.~Venugopalan, in: {In *Hwa, R.C. (ed.) et al.: Quark gluon plasma*
  249-3363}, 2003.

\bibitem{JalilianMarian:2005jf}
J.~Jalilian-Marian, Y.~V. Kovchegov, Prog. Part. Nucl. Phys. 56 (2006)
  104--231.

\bibitem{Gelis:2010nm}
F.~Gelis, E.~Iancu, J.~Jalilian-Marian, R.~Venugopalan, Ann. Rev. Nucl. Part.
  Sci. 60 (2010) 463--489.

\bibitem{Gelis:2002ki}
F.~Gelis, J.~Jalilian-Marian, Phys. Rev. D66 (2002) 014021.

\bibitem{JalilianMarian:2012bd}
J.~Jalilian-Marian, A.~H. Rezaeian, Phys. Rev. D86 (2012) 034016.

\bibitem{Rezaeian:2012wa}
A.~H. Rezaeian, Phys. Rev. D86 (2012) 094016.

\bibitem{Rezaeian:2016szi}
A.~H. Rezaeian, Phys. Rev. D93~(9) (2016) 094030.

\bibitem{Benic:2016yqt}
S.~Benic, K.~Fukushima, Nucl. Phys. A958 (2017) 1--24.

\bibitem{Benic:2016uku}
S.~Benic, K.~Fukushima, O.~Garcia-Montero, R.~Venugopalan (2016).
\newblock \href {http://arxiv.org/abs/1609.09424} {\path{arXiv:1609.09424}}.

\bibitem{Blaizot:2004wu}
J.~P. Blaizot, F.~Gelis, R.~Venugopalan, Nucl. Phys. A743 (2004) 13--56.

\bibitem{Gelis:2005pt}
F.~Gelis, Y.~Mehtar-Tani, Phys. Rev. D73 (2006) 034019.

\bibitem{Fukushima:2008ya}
K.~Fukushima, Y.~Hidaka, Nucl. Phys. A813 (2008) 171--197.

\bibitem{McLerran:1998nk}
L.~D. McLerran, R.~Venugopalan, Phys. Rev. D59 (1999) 094002.

\bibitem{Dominguez:2011wm}
F.~Dominguez, C.~Marquet, B.-W. Xiao, F.~Yuan, Phys. Rev. D83 (2011) 105005.

\bibitem{Garcia-Montero}
O.~Garcia-Montero, Master's Thesis, Ruprecht-Karls-Universit{\"{a}}t Heidelberg
  (2016).

\bibitem{Blaizot:2004wv}
J.~P. Blaizot, F.~Gelis, R.~Venugopalan, Nucl. Phys. A743 (2004) 57--91.

\bibitem{McLerran:1993ni}
L.~D. McLerran, R.~Venugopalan, Phys. Rev. D49 (1994) 2233--2241.

\bibitem{McLerran:1993ka}
L.~D. McLerran, R.~Venugopalan, Phys. Rev. D49 (1994) 3352--3355.

\bibitem{McLerran:1994vd}
L.~D. McLerran, R.~Venugopalan, Phys. Rev. D50 (1994) 2225--2233.

\bibitem{Fujii:2006ab}
H.~Fujii, F.~Gelis, R.~Venugopalan, Nucl. Phys. A780 (2006) 146--174.

\bibitem{Fujii:2013gxa}
H.~Fujii, K.~Watanabe, Nucl. Phys. A915 (2013) 1--23.

\bibitem{Kang:2013hta}
Z.-B. Kang, Y.-Q. Ma, R.~Venugopalan, JHEP 01 (2014) 056.

\bibitem{Ma:2015sia}
Y.-Q. Ma, R.~Venugopalan, H.-F. Zhang, Phys. Rev. D92 (2015) 071901.

\bibitem{Ma:2014mri}
Y.-Q. Ma, R.~Venugopalan, Phys. Rev. Lett. 113~(19) (2014) 192301.

\bibitem{Belogianni:2002ic}
A.~Belogianni, et~al., Phys. Lett. B548 (2002) 129--139.

\bibitem{Fukushima:2007dy}
K.~Fukushima, Y.~Hidaka, JHEP 06 (2007) 040.

\bibitem{Balitsky:1978ic}
I.~I. Balitsky, L.~N. Lipatov, Sov. J. Nucl. Phys. 28 (1978) 822--829, [Yad.
  Fiz.28,1597(1978)].

\end{thebibliography}







\end{document}